\documentclass[nonatbib,3p,twocolumn]{elsarticle}

\usepackage[utf8]{inputenc}           %
\usepackage[T1]{fontenc,url}
\makeatletter
\let\c@author\relax
\makeatother

\usepackage{balance}

\usepackage[abbreviate=true, dateabbrev=true, isbn=false, doi=false, urldate=comp, url=false, maxbibnames=9, backref=false, backend=biber, style=numeric-comp, language=american, giveninits=true, sortcites=true, sorting=none]{biblatex}
\usepackage{hyperref}
\usepackage{amsmath} %
\usepackage{graphicx}
\usepackage{xspace}
\usepackage{algorithm2e, etoolbox}
\AtBeginEnvironment{algorithm}{\small}
\usepackage{booktabs}
\usepackage{subcaption} %
\usepackage{ifthen}
\usepackage{fancyhdr}
\usepackage{ccicons}

\hypersetup{hidelinks}
\DeclareGraphicsExtensions{.pdf,.jpg,.png}

\let\orig\textsf
\let\sffontsize\small
\renewcommand{\textsf}[1]{\orig{\sffontsize #1}}   %
\newcommand{\tablefont}{\footnotesize\let\sffontsize\footnotesize}
\newcommand{\sfs}[1]{\textsf{#1}}
\newcommand{\threshold}{\textsf{threshold}\xspace}
\newcommand{\K}{\textsf{cutoff}\xspace}
\newcommand{\cutoff}{\textsf{cutoff}\xspace}
\newcommand{\TP}{\textit{TP}\xspace}
\newcommand{\FP}{\textit{FP}\xspace}
\newcommand{\TN}{\textit{TN}\xspace}
\newcommand{\FN}{\textit{FN}\xspace}
\newcommand{\Fi}{\textit{F$_1$}\xspace}
\newcommand{\Fii}{\textit{F$_2$}\xspace}
\newcommand{\vdisc}{\textsf{VDISC}\xspace}
\newcommand{\CWE}{\vdisc}
\newcommand{\loo}{\textsf{loo}\xspace}
\newcommand{\win}{\textsf{within}\xspace}
\newcommand{\cz}{\textsf{F-0}\xspace}
\newcommand{\call}{\textsf{F-all}\xspace}
\newcommand{\LAVDNN}{LAVDNN\xspace}
\newcommand{\FAVD}{FAVD\xspace}
\newcommand{\FAVDx}[1]{FAVD$_{\mathit{#1}}$}    %
\newcommand{\FAVDL}{\FAVDx{L}\xspace}
\newcommand{\FAVDF}{\FAVDx{F}\xspace}
\newcommand{\Rank}{\textsf{Rank}\xspace}
\newcommand{\amodel}{{$\mathcal{R}_{\mathit{L}}$}\xspace} 
\newcommand{\acount}{{$\mathcal{R}_{\mathit{F}}$}\xspace}
\newcommand{\strawman}{\emph{all-vulnerable}\xspace}

\newcommand{\head}[1]{\par\noindent\textbf{#1}}
\newcommand{\figref}[2][]{Figure~\ref{#2}#1}  %

\newtheorem{definition}{Definition}

\SetFuncSty{textsf}
\SetKwFunction{Split}{Split}
\SetKwFunction{Rank}{Rank} %
\SetKwFunction{Find}{FindBest} %
\SetKwFunction{Compute}{F$_2$}
\SetKwFunction{Unique}{Unique}
\SetKwFunction{Update}{UpdateCounts}
\SetKwFunction{Split}{Split}
\SetKwInOut{Input}{input}
\SetKwInOut{Output}{output}
\SetKwComment{Comment}{$\triangleright$~}{}

\title{Featherweight Assisted Vulnerability Discovery}
\author{David Binkley}
\address{Loyola University Maryland, 4501 N. Charles St., Baltimore, MD 21210-2699, USA}
\author{Leon Moonen}
\address{Simula Research Laboratory, Oslo, Norway}
\author{Sibren Isaacman}
\address{Loyola University Maryland, 4501 N. Charles St., Baltimore, MD 21210-2699, USA}

\fancypagestyle{plain}{%
\fancyfoot[LO RO]{}
\fancyfoot[CO]{\small%
  \raisebox{1cm}[0pt]{\parbox{\textwidth}{\centering Accepted for publication in Information and Software Technology,
  \textsc{doi}:~\href{https://doi.org/10.1016/j.infsof.2022.106844}{10.1016/j.infsof.2022.106844}.\\[2ex]
  \includegraphics[width=2.3cm]{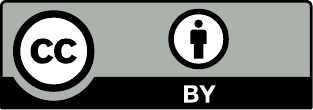}\hspace*{2mm}\raisebox{3mm}{\parbox{7.3cm}{This work is licensed under a 
  \href{https://creativecommons.org/licenses/by/4.0/}{Creative Commons Attribution 4.0 International (CC BY 4.0)} license.}}}}
}
}

\begin{document}

\begin{abstract}

Predicting vulnerable source code helps to focus the attention of a developer, 
or a program analysis technique, 
on those parts of the code that need to be examined with more scrutiny.
Recent work proposed the use of function names as semantic cues 
that can be learned by a deep neural network (DNN) to aid in the hunt for
vulnerability of functions.  

Combining identifier splitting, which we use to split each function name into
its constituent words, with a novel frequency-based algorithm, we explore the
extent to which the words that make up a function's name can be used to predict 
potentially vulnerable functions.
In contrast to the \emph{lightweight} prediction provided by a DNN considering
only function names, avoiding the need for a DNN provides \emph{featherweight}
prediction.
The underlying idea is that function names that contain certain ``dangerous''
words are more likely to accompany vulnerable functions. 
Of course, this assumes that the frequency-based algorithm can be properly
tuned to focus on truly dangerous words.

Because it is more transparent than a DNN, which behaves as a ``black box'' 
and thus provides no insight into the rationalization underlying its decisions, 
the frequency-based algorithm enables us to investigate the inner workings of
the DNN.
If successful, this investigation into what the DNN does and does not learn
will help us train more effective future models.

We empirically evaluate our approach on a heterogeneous dataset containing over
73\,000 functions labeled vulnerable, and over 950\,000 functions labeled benign.
Our analysis shows that words alone account for a significant portion of the DNN's
classification ability.
We also find that words are of greatest value in the datasets with a more homogeneous vocabulary.
Thus, when working within the scope of a given project, where the vocabulary is
unavoidably homogeneous, our approach provides a cheaper, potentially
complementary, technique to aid in the hunt for source-code vulnerabilities.
Finally, this approach has the advantage that it is viable with orders of
magnitude less training data.

\end{abstract}
\begin{keyword}
model interpretability, 
vulnerability prediction,
identifier splitting,
source code vocabulary,
software security.
\end{keyword}
 \maketitle

\thispagestyle{plain}
\clearpage

\section{Introduction}
\label{sec:intro}

\noindent
Security vulnerabilities in source code are a key quality concern in software development.
Exploitation of vulnerabilities may cause financial damage, decrease users' trust, 
and, depending on the domain, introduce personal risks. 
This makes identifying vulnerabilities in the early stages of software development useful.
Automated software inspections have proven effective at identifying certain classes of security vulnerabilities in source code~\cite{anderson2003:tool,Bessey2010,pistoia2007:survey,kulenovic2014:survey}, 
but at the same time suffer from a considerable number of false positives~\cite{anderson2008:perspiration,austin2011:one,baca2013:improving}.
Manual code reviews, or software inspections~\cite{Fagan76}, have fewer problems with false positives, 
but suffer from the sheer volume of code that must be inspected~\cite{BBV2005,WAPSC02,PSMV98}.
Thus, methods that help focus the attention of a developer or program analysis technique on those parts 
of the code that should be examined with more scrutiny have the potential to lower false positives and overall workload.

Li et al.~recently proposed \LAVDNN as a lightweight approach that uses
function names as semantic cues that can be learned by a Deep Neural Network
(DNN)~\cite{li2019:lightweight}.
To be clear, \LAVDNN is not intended as a replacement for more involved
techniques that use a multitude of code features from various levels of code granularity, 
both for general defect prediction~\cite{nam2018:heterogeneous, jing2017:improved, wang2016:automatically}, 
and vulnerability specific methods~\cite{lin2020:software,sultana2020:using,dam2018:automatic,li2018:vuldeepecker}.
Instead, it is used to triage the code and thus assist a developer in deciding
where to manually inspect the code or apply more sophisticated techniques.

In a similar vein, the goal of our work is \emph{not} to outperform the state
of the art in defect prediction. 
Instead, we have two goals related to a complementary approach.  
First, we seek to study the viability of a novel word-frequency-based approach,
and second we aim to use this approach to provide a level of interpretability
to the \LAVDNN model.
If, after tuning, the frequency-based approach manages reasonable
performance, it then suggests future opportunities to improve vulnerability
predictors such as \LAVDNN by augmenting them with information gleaned from
what we refer to as dangerous words.
Though a ``dangerous word'' is not in and of itself a danger, the word may be a red flag, much in the same way that a code smell is not a problem in and of itself, but
suggests a point of potential concern.

If successful, the frequency-based approach provides a \emph{featherweight}
alternative that avoids the need to construct a DNN, which, among other things,
finds the approach viable using orders of magnitude less training data.

The paper also investigates the degree to which \LAVDNN is leveraging the
presence of dangerous words.
If the frequency approach provides similar performance, then it suggests that
we have succeeded in interpreting the learning of the DNN as we have evidence
that \LAVDNN learns to identify dangerous words.
On the other hand, a difference in performance indicates that \LAVDNN learns
something orthogonal to the words.
Thus a contribution of our work includes its exploration of model
interpretability. 
For example, although Li et al.\ claim impressive results 
(with $\Fii$-scores reaching 0.910 and 0.915 for, respectively, C/C++ and Python programs), 
our experiments, as well as the data presented in their work~\cite[Table 11]{li2019:lightweight},
show a steep drop-off in efficacy on real-world systems. 
Nevertheless, the impressive published performance of \LAVDNN both raises the
question of ``what makes it tick'' and also makes it a prime candidate for further study.
Understanding \emph{why} and \emph{when} \LAVDNN is successful may have
implications for understanding how to better discover vulnerable functions universally. 
It is therefore instructive to try and understand \emph{what} \LAVDNN is actually learning, 
either to improve \LAVDNN itself, or to develop complementary techniques.
For example, are certain words, abbreviations, or other language patterns indicative of vulnerabilities?

\head{Contributions:} 
We investigate how the individual words that make up each function name affect vulnerability predictions:
\begin{itemize}
\item We present a featherweight approach, \FAVD (Featherweight Assisted Vulnerability Discovery), 
that uses the notion of \emph{dangerous words} as semantic cues. 
The underlying idea is that when developers choose semantically sensible names, 
a vulnerable function is more likely to be given a name that contains dangerous words.

\item 
We explore two methods for identifying dangerous word:
the first, \FAVDL, uses \LAVDNN~\cite{li2019:lightweight} as classifier to
determine if a word should be considered dangerous.
\FAVDL's performance provides insights into the role that dangerous words play
in the \LAVDNN model.
The second method for identifying dangerous words, \FAVDF, identifies dangerous
words based on the \emph{frequency} of the words in the names of known
vulnerable and benign functions.
 
Comparing these two with \LAVDNN enables us to provide  
insight into what is learned by the otherwise black-box approach used by \LAVDNN.
To begin with, the comparison of \LAVDNN and \FAVDL tells us about the use of words by \LAVDNN,
but tells us nothing about the absolute value of those words in the prediction.
\FAVDF provides that baseline.
For example, if \FAVDF performs worse than \FAVDL then we can assert that
\LAVDNN is making use of features beyond words.

\item 
We empirically evaluate the predictive ability of \FAVDL and \FAVDF
using nine datasets ranging in vocabulary diversity.
Our analysis shows that words alone account for a significant portion of the
DNN's classification ability especially with more homogeneous vocabularies. 
Hence, it is feasible to train a featherweight ``triage predictor'' 
using the function names associated with past vulnerabilities of a mature
project to gain an initial focus.
Furthermore, this technique requires orders of magnitude less training data,
and can thus easily complement existing techniques for vulnerability prediction.

\end{itemize}

The paper is organized as follows: 
Sections~\ref{sec:background} and~\ref{sec:approach} present the background and the approach itself. 
Sections~\ref{sec:RQs} and~\ref{sec:design} introduce our research questions and experimental design, 
followed by a discussion of results in Section~\ref{sec:results}.
We survey related work in Section~\ref{sec:relwork} and conclude in Section~\ref{sec:conc}.

\section{Background}
\label{sec:background}

\subsection{\LAVDNN}

\noindent
The model \LAVDNN was trained on an (undisclosed\footnote{~We have requested this data from the authors but could not obtain it.\label{fn:data}}) dataset of 
8\,525 vulnerable function names extracted from the Common Vulnerabilities and Exposures (CVE) database, 
and 8\,000 benign function names extracted from open-source projects~\cite{li2019:lightweight}. 
Each name was one-hot encoded into a matrix of 66 rows (for the allowed alphanumeric characters) 
and 50 columns (for the allowed maximum function-name length),
which together with its label as benign or vulnerable, 
was used to train a multi-layer Bidirectional Long Short Term Memory (LSTM) network for classification.
\LAVDNN concludes with two densely connected output nodes whose values are run through a softmax function.
Thus, the network outputs the likelihoods that a function is ``vulnerable'' or ``benign''
as values in the range 0.0--1.0.
The authors experimentally determine that a
threshold value of 0.55 for the ``vulnerable'' class provides the best performance.
Thus, function names with a score of 0.55 or greater in the ``vulnerable'' output node are classified as vulnerable
and the rest benign.

Care must be taken when referring to the paper that introduced
\LAVDNN~\cite{li2019:lightweight}.
While the paper claims very impressive $\Fii$ values,
computing $\Fii$ values using data from the paper's Table 11 produces much lower $\Fii$ values for real-world systems (e.g., 0.683 for \sfs{LibTIFF} and 0.746 for \sfs{FFmpeg}).
These values cannot be reproduced, because the paper's limited replication package, 
which only includes the function names from these two systems, 
does not label them as \emph{vulnerable} or \emph{benign}.
Fortunately, Lin et al.~\cite{lin2018:crossproject} independently provide
the necessary data, albeit for slightly newer versions of \sfs{LibTIFF} and \sfs{FFmpeg}.
However, as shown in the last column of Table~\ref{tab:all-vul} discussed later,
applying \LAVDNN to this data results in the
notably lower $\Fii$ scores of 0.292 for \sfs{LibTIFF} and 0.083 for \sfs{FFmpeg}.

\subsection{Identifier Splitting}

\noindent
Identifier splitting~\cite{caprile1999:nomen} splits an identifier into its
constituent parts, called \emph{terms}.
For example, the identifier \sfs{read\_file} includes the terms
\sfs{read} and \sfs{file}.
Splitting algorithms range from conservative (looking for Camel and
Snake case) to aggressive (e.g.,~able to separate \textsf{maxstrlen} into
\textsf{max}, \textsf{str}, and \textsf{len})~\cite{hill2014:empirical}.

\section{Approach}
\label{sec:approach}

\noindent
This section describes \FAVD, our algorithm for \emph{featherweight assisted
vulnerability discovery}, the core of which is given as Algorithm~\ref{alg:favd}.
{The algorithm takes as input three parameters and outputs the set of
identifiers predicted to be vulnerable, $\mathcal{V}$.
The output is a subset of the first input parameter, $\mathcal{I}$, which is
the set of identifiers being tested.  
The second input parameter is the training data, $\mathcal{T}$, which is a set
of identifiers each labeled as \emph{vulnerable} if it is the name of a
function with a vulnerability and \emph{benign} otherwise.
The final input parameter, \sfs{min\_score}, is the minimum score that an
identifier must receive to be returned by \Rank as a dangerous word.
}
The algorithm first conservatively splits each function name from the training
data into its constituent terms.
The resulting set of terms is used as the source of potential dangerous words:

\begin{definition}
A word is a \emph{dangerous word} if its presence in a function's name
correlates with the function being more likely to include a vulnerability.
\end{definition}

\noindent
For example, functions that accept user input are often vulnerable to various
stack attacks.
The names of such functions often include words such as \emph{read} or
\emph{input}. Thus, we may classify \emph{read} and \emph{input} as ''dangerous,'' marking functions using those terms as potentially vulnerable.

\FAVD's primary goal is converting a set of terms into a ranked list of
dangerous words.
This is done by the function \Rank, which takes the set of terms and a minimum
(dangerousness) score.
This function first assigns a dangerousness score to each term and then
discards those terms whose score is less than the given minimum, \sfs{min\_score}.
It returns a list of the remaining terms in decreasing order of dangerousness.

\begin{algorithm}[b]
\small
\DontPrintSemicolon
\SetAlgoLined
\Input{test identifiers $\mathcal{I}$, labeled training data $\mathcal{T}$, \sfs{min\_score}}
\Output{set of vulnerable identifiers, $\mathcal{V}$}
	\sfs{terms} $\leftarrow$ \Unique(\Split($\mathcal{T}$)) \\
    dangerous $\leftarrow$ \Rank(terms, min\_score)) \\
   	cutoff, threshold $\leftarrow$ \Find(dangerous, $\mathcal{T}$) \\
   	\For {\sfs{id} $\in \mathcal{I}$}{
      terms $\leftarrow$  \Unique(\Split(id)) \\
      percentage $\leftarrow$ $|$terms $\cap$ dangerous[1..cutoff]$| / |$terms$|$ \\
      $\mathcal{V}$ $\leftarrow$ $\mathcal{V}$ $\cup$ \{ id \} if percentage $>$ threshold \\
  }
\medskip
\caption{\textrm{\footnotesize\FAVD vulnerability prediction\label{alg:favd}.}}
\end{algorithm}

\FAVD next calls the function \sfs{FindBest}, which takes the list of dangerous
words and the training data, and outputs two values: \K and \threshold.
The \K determines how many of the dangerous words are retained when predicting the
vulnerability of the test data found in $\mathcal{I}$.
If the percentage (see Algorithm~\ref{alg:favd}) of a function-name's terms
that are dangerous is greater than \threshold, 
then the function is predicted to be vulnerable.

\sfs{FindBest} searches for a \emph{winning combination} of these two.
For example, having a small \K means a short list of dangerous words, 
which often works better with a low \threshold, 
because with few dangerous words,
most identifiers will have at most a few dangerous terms.
On the other hand, when \K is large, 
there are lots of dangerous words, 
and a higher \threshold often works better.

As mentioned in Section~\ref{sec:intro}, we experiment with two different
vulnerability discovery algorithms, \FAVDL and \FAVDF.
These two differ only in the implementation of the function \Rank.
In the ideal case, \Rank returns exactly those words that will identify the
vulnerable identifiers in the test data, $\mathcal{I}$.
We approximate this ideal using two different ranking functions, \amodel and \acount. 
The first, \amodel, uses \LAVDNN to determine each term's
dangerousness. 
Here, each term found in a function name is fed into \LAVDNN in isolation, 
thereby producing a score for the term between 0.00 and 1.00.

The second ranking function, \acount, is based on term frequency, and is thus
entirely independent of \LAVDNN.
This enables us to better understand the value that the words that make up a
function's name play in \LAVDNN's assessment.
It produces integer term scores as follows:
for each vulnerable identifier in the training data it increments the
dangerousness score of all of the identifier's terms by the constant
\sfs{plus}, while for each benign identifier it decrements this
score by the constant \sfs{minus}.
Using different pairs of constants, referred to as \emph{weights}, allows
\acount to place more or less emphasis on terms frequently found in the
vulnerable or benign training data.

\section{Research Questions}
\label{sec:RQs}

\begin{description}
\item [RQ1]
\emph{What is the result of being excessively conservative and declaring all
functions vulnerable?} --
While {simplistic}, this is the lightest-weight of approaches and
is guaranteed to have high recall, so it sets a good baseline.

\item [RQ2]
\emph{Can \LAVDNN be used to construct a list of dangerous words that can effectively predict vulnerable functions?} --
In other words, can \LAVDNN predict which terms are associated with vulnerable functions?

\item [RQ3]
\emph{Does direct construction of a list of dangerous words provide insight to what \LAVDNN learns?} --
In order to investigate the potential value that terms might bring to the DNN, 
we use \emph{term frequency} as an alternative method of determining the list of
dangerous words.

\end{description}

\newpage

\section{Experimental Design}
\label{sec:design}

\subsection{Datasets and Ground Truth}

\noindent
We consider the nine datasets shown in Table~\ref{tab:data-sets} 
(note that the ninth dataset, \win, aggregates the first six datasets).
For each dataset, we work exclusively with lists of benign and
vulnerable function names.
We clean each list by removing ``internal'' duplicates 
(caused when two or more functions share the same name).
For each dataset, we make the lists of benign and vulnerable names disjoint,
taking the conservative stance that names appearing on both 
lists are potentially vulnerable. 
A replication package with our data is available on GitHub and 
Zenodo.\footnote{~\url{https://github.com/secureIT-project/FAVD}\newline
\hspace*{6.5mm}\textsc{doi:}~\href{https://doi.org/10.5281/zenodo.5957264}{10.5281/zenodo.5957264}}

\begin{table}\tablefont
\caption{Overview of the nine datasets used in the study. Note that the ninth dataset, \win, aggregates the first six.}
\label{tab:data-sets}
\centering
\raisebox{-.4em}{\raisebox{.75em}{\rotatebox{90}{\win}}\,\,\scalebox{1.1}[7]{\{\!}}
\setlength{\tabcolsep}{5pt}
\begin{tabular}{l r r r r}
\toprule
\multicolumn{1}{l}{dataset}
  & \multicolumn{1}{c}{vulnerable}
  & \multicolumn{1}{c}{benign} 
  & \multicolumn{1}{c}{\% vuln.} 
  & \multicolumn{1}{c}{overlap} \\
\midrule
\sfs{Asterisk}  &       49  & 10\,102 &  0.5\% &   2 \\   %
\sfs{FFmpeg}    &      184  &  4\,379 &  4.2\% &  18 \\
\sfs{LibPNG}    &       31  &     491 &  6.3\% &   0 \\
\sfs{LibTIFF}   &       75  &     522 & 13.6\% &   8 \\
\sfs{Pidgin }   &       26  &  6\,722 &  0.3\% &   0 \\
\sfs{VLC}       &      37 &   2\,699  &  1.4\% &   3 \\
\midrule
\loo            &     402 &  24\,906  &  1.6\% &  33 \\
\midrule
\CWE            & 72\,612 & 932\,741  &  7.2\% & 11\,970 \\[2pt]
\bottomrule
\end{tabular}
\end{table}

The first six datasets in Table~\ref{tab:data-sets} come from data shared by Lin et
al.~\cite{lin2018:crossproject}.
They extracted 457 vulnerable functions from six open-source projects based on CVE
reports~\cite{mitre:cve} and added 32\,531 benign functions from each
project's source code repository.
Table~\ref{tab:data-sets} shows the sizes for each after cleaning.
Asterisk is a C++ library for PBX integration and 
Pidgin is a library for developing chat clients. 
The other four projects are from more closely related domains, 
with FFmpeg and VLC being well-known video applications, 
and LibPNG and LibTIFF providing image manipulation. 
The main programming language for all projects is C, 
with small amounts of C++, Python, HTML, Shell, and Assembly.
Table~\ref{six-project-demographics} summarizes 
demographic details of the six projects of the \win dataset. 

We first consider these six in isolation, performing $k$-fold cross-validation
separately, per project.
That is, we use the names from each project independent of the other projects.
Given the identifiers in these experiments are coming from a single project, they are expected to have the
least \emph{diversity} in their vocabulary.
{By the \emph{diversity} of a vocabulary, we mean the variety of words used.
For example, the vocabulary built from the identifiers 
\sfs{remove\_node} and \sfs{remove\_edge} show less
diversity than the one built from the identifiers
\sfs{remove\_node} and \sfs{delete\_edge}.}
This gives us an indication of how well a proposed algorithm performs when
applied to a mature system, 
where there exists vulnerability data from exploits found in older versions of the system to train against.
When comparing these six to the \loo and \CWE datasets, we often aggregate them
by taking means.  We refer to this aggregate as the \win dataset.

 \newcommand{\tpt}{\textperthousand}
 \begin{table*}[t]\tablefont
 	\centering
 	\caption{Demographic details characterizing the projects in our dataset (src: OpenHub~\cite{synopsys:openhub})}
 	\label{six-project-demographics}
 	\begin{tabular}{lrrrrrr}
		\toprule
 		features                    & \sfs{Asterisk} & \sfs{FFmpeg} & \sfs{LibPNG} & \sfs{LibTIFF} & \sfs{Pidgin} & \sfs{VLC} \\
 		\midrule
		\#contributors              &            302 &       1\,968 &           58 &            64 &          790 &       924 \\ 
 		total LOC                   &        2\,529k &      1\,248k &         462k &          267k &         224k &      717k \\ 
        estimated effort (years)    &            741 &          356 &          122 &            68 &           52 &       197 \\
		\#commits                   &        96\,265 &     102\,835 &      10\,679 &        6\,447 &      40\,605 &   88\,567 \\
		files modified              &        17\,777 &      10\,082 &       4\,724 &        1\,830 &      12\,199 &   16\,012 \\
		lines added                 &       14\,051k &      4\,292k & 	   2\,378k &       1\,229k &      8\,321k &   4\,989k \\
        lines removed               &        9\,996k &      2\,316k &      1\,696k &          938k &      7\,549k &   3\,991k \\
 		security confidence         &        97.84\% &      95.14\% &      85.41\% &       91.35\% &         N/A* &      N/A* \\ 
 		vulnerability exposure      &        1.1\tpt &      4.3\tpt &     27.0\tpt &      21.8\tpt &         N/A* &      N/A* \\ 
 		\bottomrule
		\multicolumn{7}{c}{* there were no vulnerabilities reported for \sfs{Pidgin} and \sfs{VLC} in OpenHub}   
 	\end{tabular}
 \end{table*}

The next dataset, \loo, makes use of the same six open-source projects.
However, this time we perform \emph{leave-one-out} cross-validation on the set
of six.
Leave-one-out cross-validation uses the vulnerability of function names from
all but one of the projects to predict the vulnerability of function names from
the one left out.
The names are not expected to be as similar as in the first six datasets (i.e., they have higher diversity), but because
a number of the projects are from similar domains, we expect some similarity.

The largest dataset, \CWE, was extracted from the data published by Russell et al.~\cite{russell2018:automated}.
It contains 1.27 million functions mined from open-source software, 
labeled for potential vulnerability by static analysis tools. 
After cleaning, we end up with 1\,005\,353 function names,
including 72\,612 marked as vulnerable and 932\,741 marked as benign. 
Because some of this data is likely included in the \win and \loo datasets,
we never combine it with either of the two, however studying them separately
improves the external validity of our analysis.  
For example, one interesting difference is that the percentage of vulnerable
identifiers in the \loo dataset is only 1.6\%, which is a notably smaller than
the 7.2\% of the \CWE dataset.

\subsection{Performance Measures} 

\noindent
Our goal is to classify function names as either vulnerable or benign.
We define true positives, $\TP$, as the correctly identified
vulnerable functions, true negatives, $\TN$, as the correctly identified benign
functions, false positives, $\FP$, as any benign function identified as
vulnerable, and false negatives, $\FN$, as any vulnerable function identified as
benign.

The evaluation of the quality of this classification is based on a combination
of precision and recall.
\emph{Precision} is the fraction of function names determined to be vulnerable
that actually are.
In other words, $\TP/ (\TP + \FP)$.
\emph{Recall} is the fraction of all vulnerable functions correctly determined
to be vulnerable.
In other words, $\TP / (\TP + \FN)$.

Precision and recall often oppose each other.
For example, high precision is often possible by choosing only those
cases that you are very sure of, but this necessarily lowers recall. 
The balanced F-score, $\Fi$, is the mean of precision and recall, 
and thus provides a balanced combination of the two.

However, as Li et al.~observe, ``in vulnerability detecting
systems, it is first necessary to detect as many vulnerabilities as possible.
When analyzing the source code, the false reporting may increase
the workload, but failing to identify a vulnerable function
is costly and unacceptable''~\cite{li2019:lightweight}.
To support this position, they use the $\Fii$-score, which values recall over precision.
In general

\vspace*{-0.25em}{\small\begin{eqnarray*} %
F_\beta  & = & \frac{(1+\beta^2) \, Precision \times Recall}{\beta^2 \,  Precision + Recall} \\
         & = & \frac{(1+\beta^2) \TP}{(1+\beta^2) \TP + \beta^2\FN + \FP}
\end{eqnarray*}}%

\noindent
For consistency, we follow Li et al.~and use $\beta = 2$, and
thus focus the evaluation on the $\Fii$ score, $5 \TP / ( 5 \TP + 4 \FN + \FP$).

\subsection{Procedure}

\noindent
The surface goal of both \amodel and \acount is to retain only high-impact dangerous words.
However, too high a minimum (too high a value of {\sfs min\_score} in
Algorithm~\ref{alg:favd}) can starve the algorithm of sufficient vocabulary.
Therefore, we consider a range of minimum scores in the experiments.
For \amodel, we initially consider values between 0.00 and 1.00 stepping by 0.05.
We later add a few additional values to zoom in on points of interest.
For \acount, we consider only two minimums: none, where all terms are
considered dangerous, and zero, which eliminates terms of low dangerousness.
This is sufficient for \acount, because we can use the relative values of
\sfs{plus} and \sfs{minus} to impact the number of terms that receive
a positive score.
Note that when including all words on the list of dangerous words, their
relative position is still impacted by the values of \sfs{plus} and
\sfs{minus}.

Beneath the surface, we are interested in forming a better understanding of the
role that the terms found in function names play in \LAVDNN's ability to
identify dangerous functions.
Thus, while the results include some direct comparisons, we are more
interested in the deeper understanding that the relative performance and the
relative value of dangerous words bring to the prediction.

\begin{table*}\tablefont
\caption{Examples of the Most and Least Dangerous Words}   %
\label{tab:examples}
\centering
\begin{tabular}{l @{~~~~~~} l @{~~~~~~} l @{~~~~~~} l @{~~~~~~} l @{~~~~~~} l}
\toprule
  \multicolumn{1}{l}{\sfs{Asterisk}}
  & \multicolumn{1}{l}{\sfs{FFmpeg}}
  & \multicolumn{1}{l}{\sfs{LibPNG}}
  & \multicolumn{1}{l}{\sfs{LibTIFF}}
  & \multicolumn{1}{l}{\sfs{Pidgin}}
  & \multicolumn{1}{l}{\sfs{VLC}}\\
\multicolumn{4}{l}{\emph{Most Dangerous}} \\
~~invite       & avi         &  png        & JPEG             & mxit        & MP            \\
~~retrans      & 264         &  handle     & pdf              & msn         & AVI           \\
~~pkt          & vp          &  CCP        & Checked          & httpconn    & Html          \\
~~unpacksms    & old         &  CAL        & readwrite        & slp         & Strip         \\
~~aocmessage   & avcodec     &  PLT        & LZWDecode        & emoticon    & Tags          \\
~~milliwatt    & ivi         &  do         & Into             & silc        & ASF           \\
~~astman       & tile        &  PLTE       & Entry            & slplink     & vcd           \\
~~sipsock      & mjpeg       &  read       & Strips           & yahoo       & skcr          \\
~~action       & hdr         &  filter     & cvt              & idn         & LOADSparse    \\
~~ha           & gif         &  chunks     & readgitimage     & untar       & Recieve       \\

\multicolumn{4}{l}{\emph{Least Dangerous}} \\
~~asn          & ff          & image       & Samples          & purple      & Get           \\
~~ast          & get         & transform   & Handler          & cb          & vlc           \\
~~254          & write       & store       & Fax              & get         & Callback      \\
~~PD           & init        & init        & Error            & pidgin      & vlclua        \\
~~PE           & frame       & standard    & Check            & set         & Control       \\
~~225          & read        & gpc         & Image            & jabber      & Set           \\
~~get          & parse       & 16          & Swab             & add         & Add           \\
~~channel      & mov         & gp          & Set              & account     & Out           \\
~~handel       & tag         & gamma       & Proc             & blist       & test          \\
~~to           & mxf         & display     & Warning          & media       & rtp           \\

\bottomrule
\end{tabular}
\end{table*}

{
To provide some intuition for the words ranked as the most dangerous and the
least dangerous, Table~\ref{tab:examples} lists the top ten examples from
each category as identified by \acount.
Over 90\% of the top ten most dangerous words occur in more vulnerable names
than benign names.
The \sfs{LibPNG} name \sfs{handle} is a classic example of the expected
behavior.  
Of the fifteen function names that include the word \sfs{handle}, ten are on
the vulnerable list.
In extreme cases all the names are vulnerable.
For example, in \sfs{Pidgin} the word \sfs{mxit} is found in six names, all vulnerable, while
the \sfs{VLC} word \sfs{AVI} is found in two names, both vulnerable.
In \sfs{LibPNG} the word \sfs{read} occurs in five vulnerable names and 19 benign names.
In all five vulnerable names (e.g.,~\sfs{png\_push\_read\_chunk}) the top scoring word
\sfs{png} also appears. 
None of the ten least dangerous words from \sfs{LibPNG} occur in a vulnerable name.
}

{
Considering the least dangerous words, most occurrences are found in benign
names.
With the exception of \sfs{FFmpeg} there are only zero to six occurrences 
of all ten words in the vulnerable names for each program.
While for \sfs{FFmpeg} the least vulnerable names occur in 97 names, each
is counter balanced by numerous benign occurrences.
For example, \sfs{frame} occurs in 33 vulnerable names, but 293 benign ones.
}

\section{Results}
\label{sec:results}

\subsection{RQ1}

\begin{table*}\tablefont
\caption{Performance of the \strawman predictor with performance of the \LAVDNN model for comparison.}
\label{tab:all-vul}
\centering
\setlength{\tabcolsep}{8pt}
\begin{tabular}{l@{\hspace{-2pt}}r@{~}r@{\hspace{-2pt}}r r r@{}r}
\toprule
  & 
  & \multicolumn{1}{r}{}
  & \multicolumn{1}{r}{empirical}
  & \multicolumn{1}{r}{theoretical} &
  & \multicolumn{1}{r}{\LAVDNN}\\
\multicolumn{1}{c}{dataset}
  & \multicolumn{1}{r}{vulnerable}
  & \multicolumn{1}{r@{\hspace{-2pt}}}{benign}
  & \multicolumn{1}{r}{$\Fii$ score}
  & \multicolumn{1}{r}{$\Fii$ score}
  & \multicolumn{1}{r}{fold} 
  & \multicolumn{1}{r}{$\Fii$ score} \\
\midrule
\sfs{Asterisk}    &        9.8 &    2$\,$020.4  &	 0.024	& 0.024 &  mean$^*$ & 0.013 \\   %
\sfs{FFmpeg}      &       36.8 &        875.8	  &  0.173	& 0.174 & mean$^*$ & 0.083 \\  %
\sfs{LibPNG}      &        6.2 &         98.2 &  0.234 & 0.240 & mean$^*$ & 0.137 \\    %
\sfs{LibTIFF}     &       15.0  &        104.4    &  0.410  & 0.418 &  mean$^*$ & 0.292 \\    %
\sfs{Pidgin}      &        5.2  & 1$\,$344.4    &  0.019  & 0.019 &  mean$^*$ & 0.019 \\    %
\sfs{VLC}         &       7.4  &        539.8    &  0.064  & 0.064 &  mean$^*$ & 0.089 \\   %
\midrule
\loo               &       49.0  &  10$\,$102.0    &  0.042  & 0.075 &  Asterisk & 0.013 \\
\loo               &      184.0  &   4$\,$379.0    &  0.274  & 0.075 &  FFmpeg & 0.083 \\
\loo               &       31.0  &        491.0    &  0.362  & 0.075 &  LibPNG & 0.137 \\
\loo               &       75.0  &        522.0    &  0.564  & 0.075 &  LibTIFF & 0.292 \\
\loo               &       26.0  &   6$\,$722.0    &  0.034  & 0.075 &  Pidgin & 0.019 \\
\loo               &       37.0  &   2$\,$699.0    &  0.110  & 0.075 &  VLC & 0.089 \\
\midrule
\CWE          &  14$\,$522.4  & 186$\,$548.2  &  0.280  & 0.280 &   mean$^*$ & 0.148 \\    %
\bottomrule
\multicolumn{7}{l@{}}{\footnotesize $^*$~For brevity, we present the mean values for 5-fold cross-validation.} \\
\multicolumn{7}{l@{}}{\footnotesize \phantom{$^*$}~10-fold cross-validation showed no statistically significant differences.}
\end{tabular}
\end{table*}

\noindent
Our first research question considers the results of being excessively
conservative by predicting that all functions are vulnerable.
This approach provides intuition for the meaningful range of $\Fii$ scores.
A perfect predictor attains the highest possible $\Fii$ score of 1.0, 
while the worst-case $\Fii$ score is 0.0.
The $\Fii$ score for a random predictor depends on the percentage of
vulnerable functions.
For example, the \CWE dataset, with its 72$\,$612 vulnerable functions and
932$\,$741 benign ones, is 7.2\% vulnerable, meaning a random predictor
will generate \TP = \FN = 3.6\% and \FP = \TN = 46.4\%, yielding an $\Fii$
score of 0.228. When only 1.6\% of the functions are vulnerable, the resulting
$\Fii$ score drops to 0.071.

Assuming that all functions are vulnerable, there are no false negatives or
true negatives, and thus \FN and \TN are both zero.
In part because $\Fii$ favors recall, the \strawman assumption will yield predicted $\Fii$
scores slightly higher than random.
For example, the \CWE dataset with its 7.2\% vulnerable functions, results in
an \strawman predicted $\Fii$ score of 0.280.

Conversely, given our imbalanced dataset, a high-accuracy strategy
is to classify all functions as being in the dominant class.
In our case, most functions are benign. %
Predicting that all functions are benign produces very high accuracies (e.g.,~0.928 for the
\CWE dataset). However, other than highlighting a potential misinterpretation,
classifying all functions as benign makes little sense in the context of
vulnerability prediction;
thus, we consider it no further.  %

In support of RQ1, Table~\ref{tab:all-vul} shows the {empirical and
theoretical} $\Fii$ values for each dataset.
These two scores provide some intuition for the performance of the simple
\strawman predictor.
The empirical $\Fii$ value for each fold is based on the number of vulnerable
and benign functions in that fold.
{For the $k$-fold validations this leads to some minor variation that we
summarize in the table using the mean.}
{For the \loo dataset the empirical values deviate from the theoretical values because the distribution of vulnerable names is far less uniform.}
Table~\ref{tab:all-vul} also shows performance of \LAVDNN for comparison.

Thus, in summary for RQ1, Table~\ref{tab:all-vul} shows that the $\Fii$ scores
for this ultra-conservative approach range from 0.034 to 0.564 in line with the
relative number of vulnerable and benign functions in each fold of each dataset. 
While its 100\% recall may be the most redeeming quality, 
the \strawman predictor sets a baseline for the
minimum performance expected of a more sophisticated predictor.  
In RQ2 and RQ3, our goal is to provide an interpretability viewpoint for the
\LAVDNN model.
We accomplish this by replacing all words being dangerous with more sophisticated
techniques that we then compared to \strawman and to each other.

\subsection{RQ2}
\label{sec:RQ2}

\begin{figure}
  \centering
  \includegraphics[width=0.99\columnwidth]{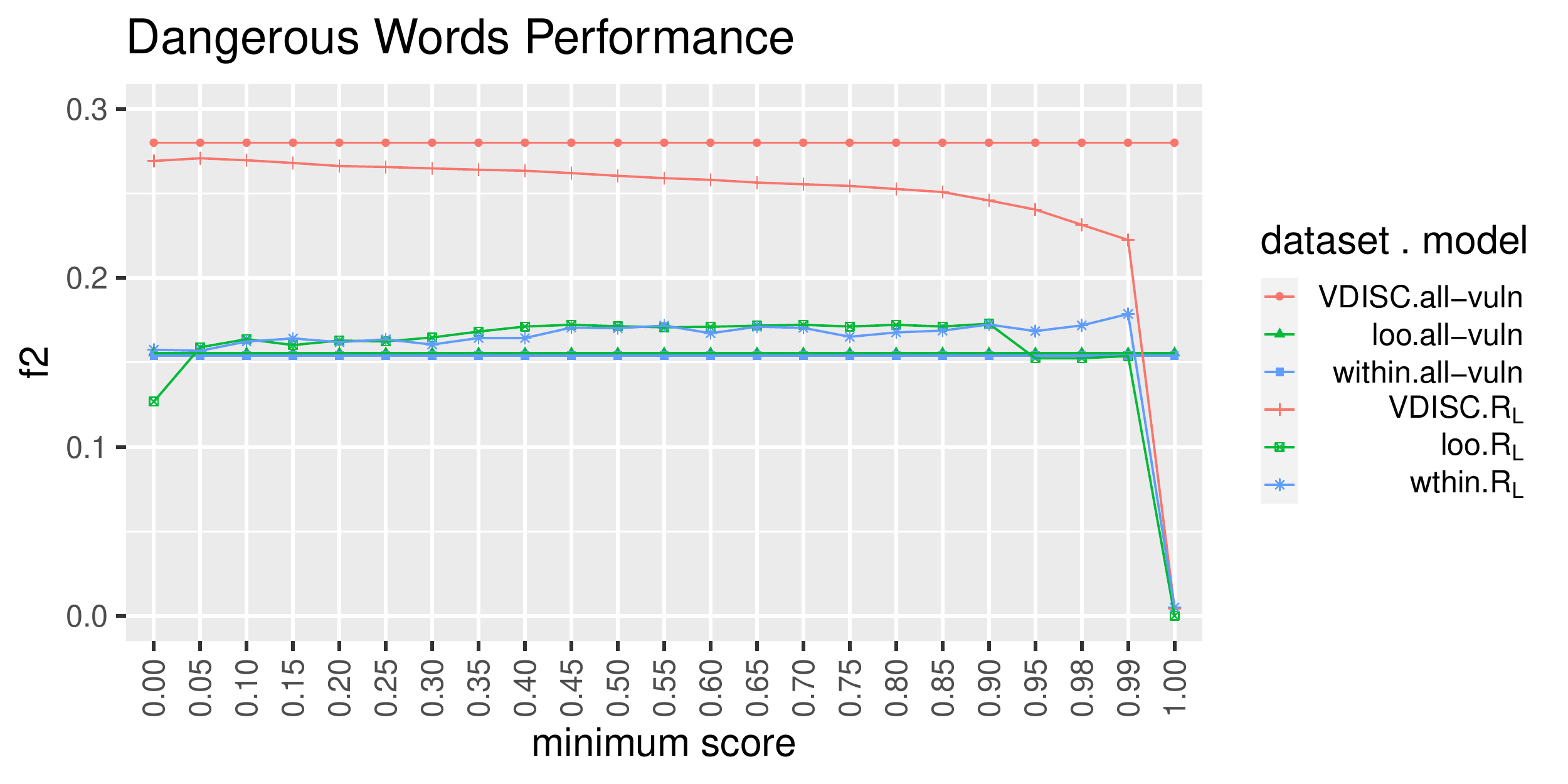}
  \includegraphics[width=0.99\columnwidth]{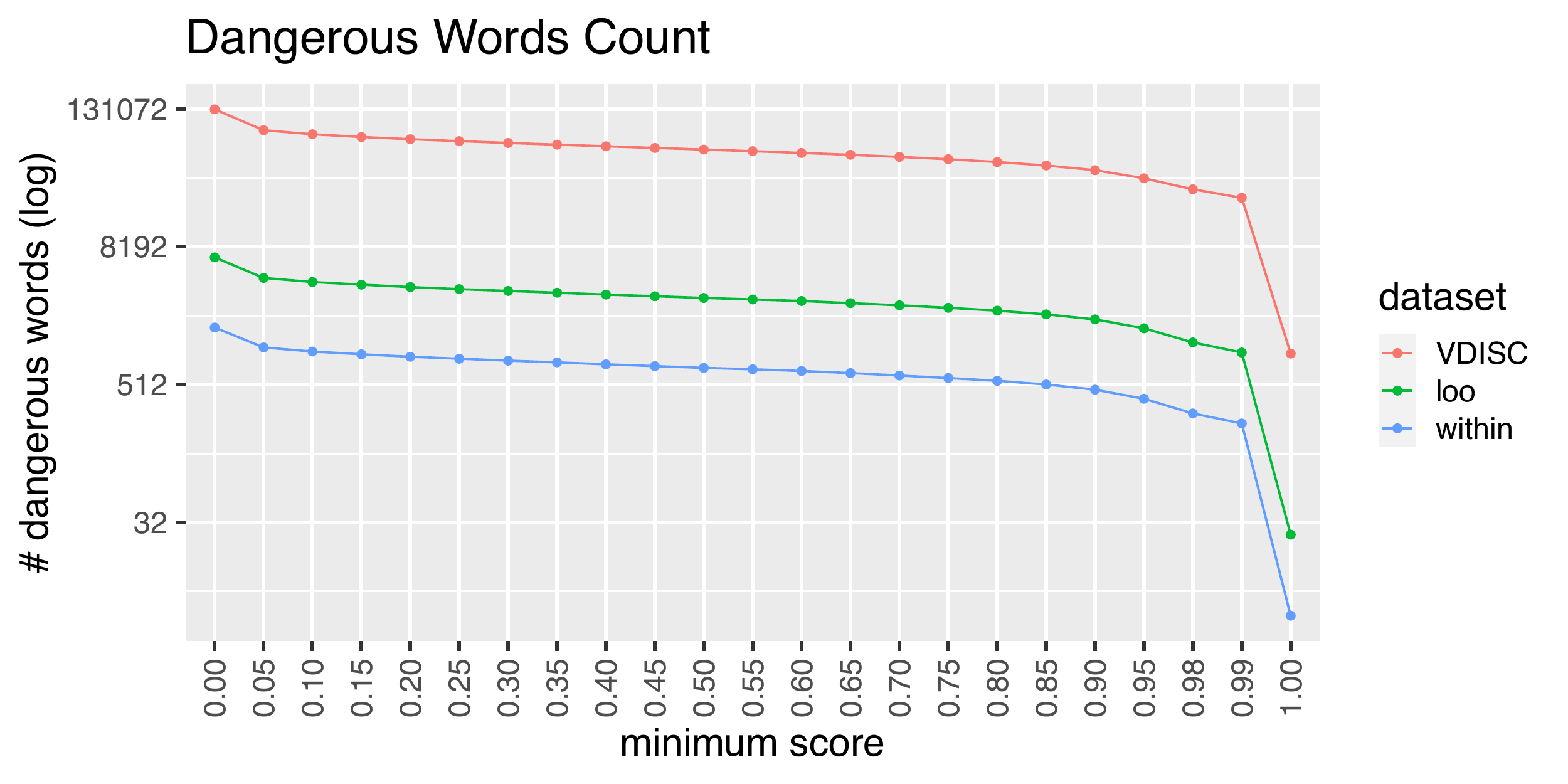}
  \caption{Using \amodel to select dangerous words.}
  \label{fig:min_score}
\end{figure}

\begin{figure*}
  \centering
  \includegraphics[width=0.99\columnwidth]{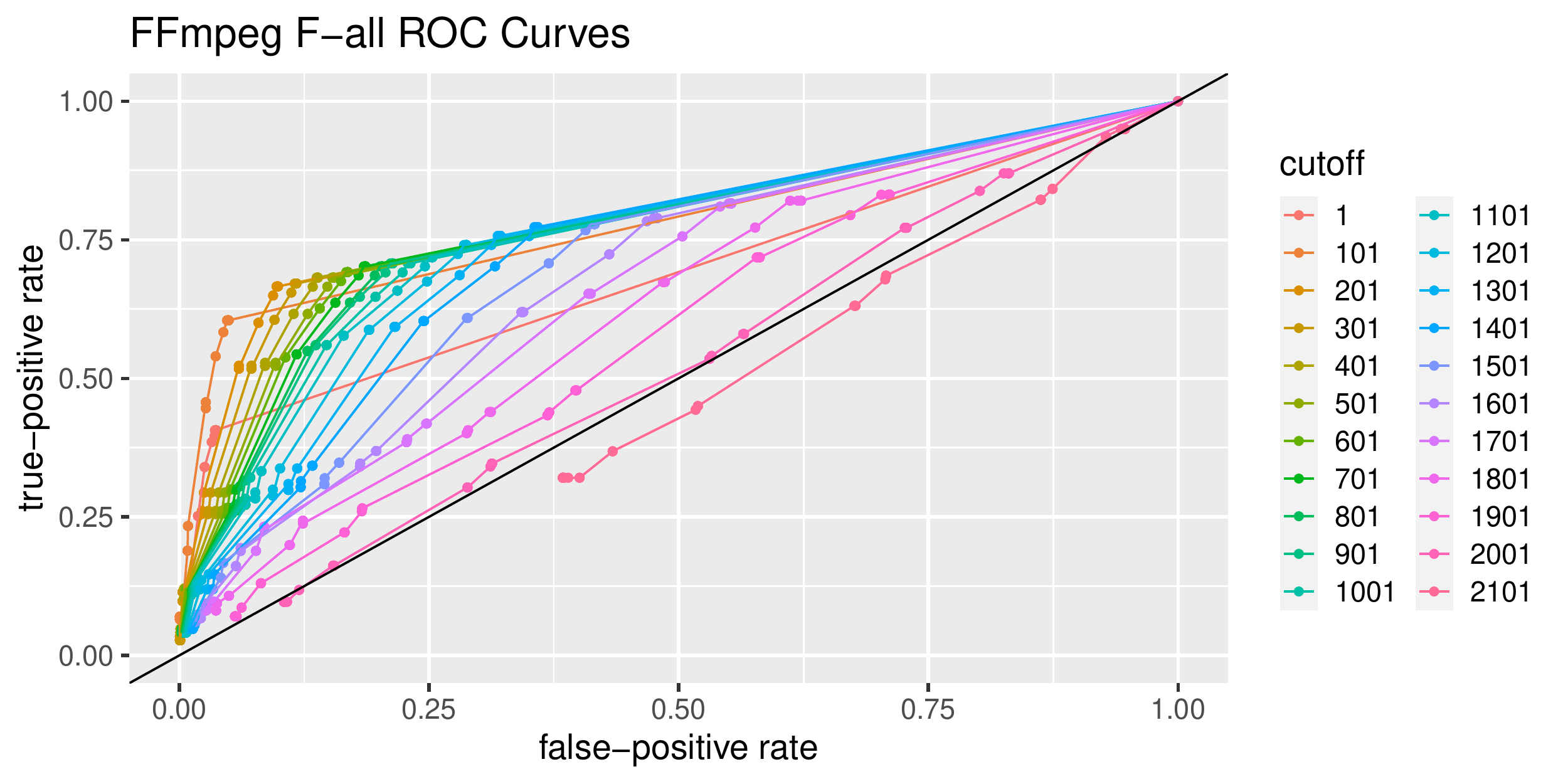} \\
  \caption{ROC curves exploring the search space for \cutoff and \threshold}
  \label{fig:ROC-FFmpeg}
\end{figure*}

\noindent
RQ2 investigates using \amodel as the \sfs{Rank} function in
Algorithm~\ref{alg:favd}, that is, it studies how well \LAVDNN can predict
\emph{dangerous words}.
We begin our analysis with the two graphs shown in Figure~\ref{fig:min_score}.
The upper graph shows the minimum score on the $x$-axis and the $\Fii$ score on the $y$-axis. 
Note that the $x$-axis is not to scale because values between 0.90 and 1.00 are more interesting.
The line colors represent the three datasets \win, \loo, and \CWE.
For comparison, the three horizontal lines show the $\Fii$ scores from the
ultra-conservative \strawman predictor considered in RQ1 when applied to the
three datasets.

The lower chart in Figure~\ref{fig:min_score} shows how increasing the required
minimum score decreases the number of words considered dangerous (variable
\sfs{dangerous} in Algorithm~\ref{alg:favd}). 
With this decrease, the expectation is that the remaining, higher scoring, words
will be better predictors, but apply to fewer vulnerable functions.
{This pattern is just evident in the very slight increase in the $\Fii$
values for the \loo and \win datasets in the top chart.  However the pattern is
not strong and the trend for \vdisc actually declines (reflecting the dataset
being starved for useful vocabulary).}
However, the fact that the $\Fii$ scores are relatively flat suggests that the
reduction in the number of dangerous words is not providing greater
discrimination, and thus that words given higher scores by \amodel are not
necessarily better predictors of vulnerable functions.

Objectively, for \CWE, \amodel always underperforms the \strawman predictor 
(top two lines in the top chart of Figure~\ref{fig:min_score}).
However, \amodel outperforms the \strawman predictor on the other two datasets.
Omitting the minimum score of 1.00, where all three $\Fii$ scores plummet to near zero, 
all three differences are statistically significant (t-test $p < 0.0001$ for
\CWE and \win, and $p = 0.0011$ for \loo, using the data shown in Figure~\ref{fig:min_score}).

In summary, \LAVDNN finds limited success at identifying dangerous words.
\amodel works better with the smaller, more focused, data sets of \win and \loo.
However, we note two caveats:
first, the data clearly show that using too high a minimum score leads to too
few dangerous words, which dramatically lowers the $\Fii$ score, and 
second, it must be pointed out that on an absolute scale, the resulting $\Fii$
scores are all on the low side.

Looking ahead to the comparison with \acount, we note that numerically the best
performance for \CWE is with a minimum score of 0.05, while for \loo it is
0.90, and for \win 0.99.
These values, which are all evident in Figure~\ref{fig:min_score}, reflect
artifacts of the vocabularies.
For example, for \CWE, finding the best performance with such a low minimum
score indicates that the search is starved for high-quality vocabulary, while
at the other end of the spectrum, for \win, the very high minimum excludes all
but the most suspect words.

\begin{figure*}
  \centering
  \includegraphics[width=0.99\columnwidth]{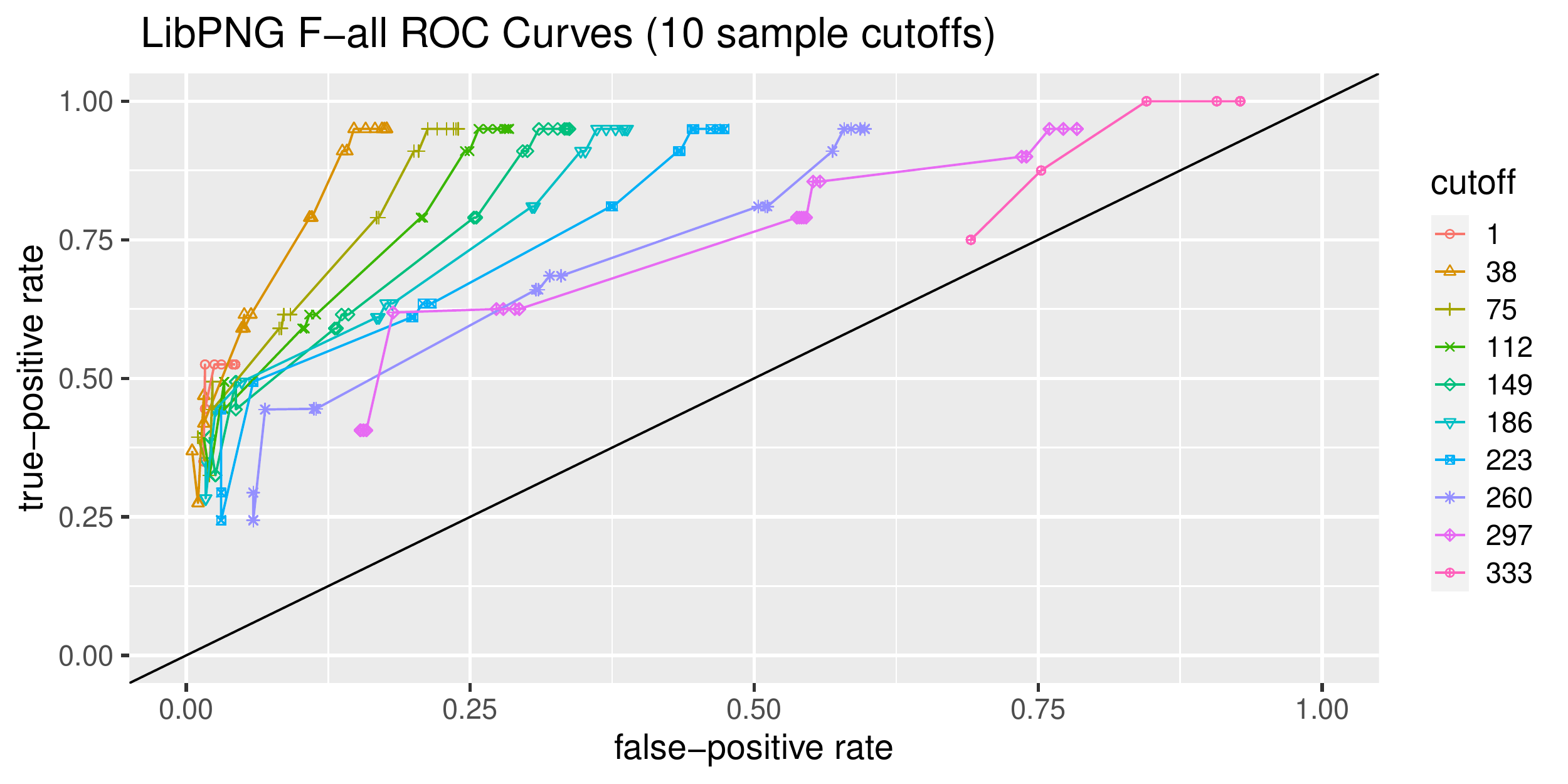} 
  \includegraphics[width=0.99\columnwidth]{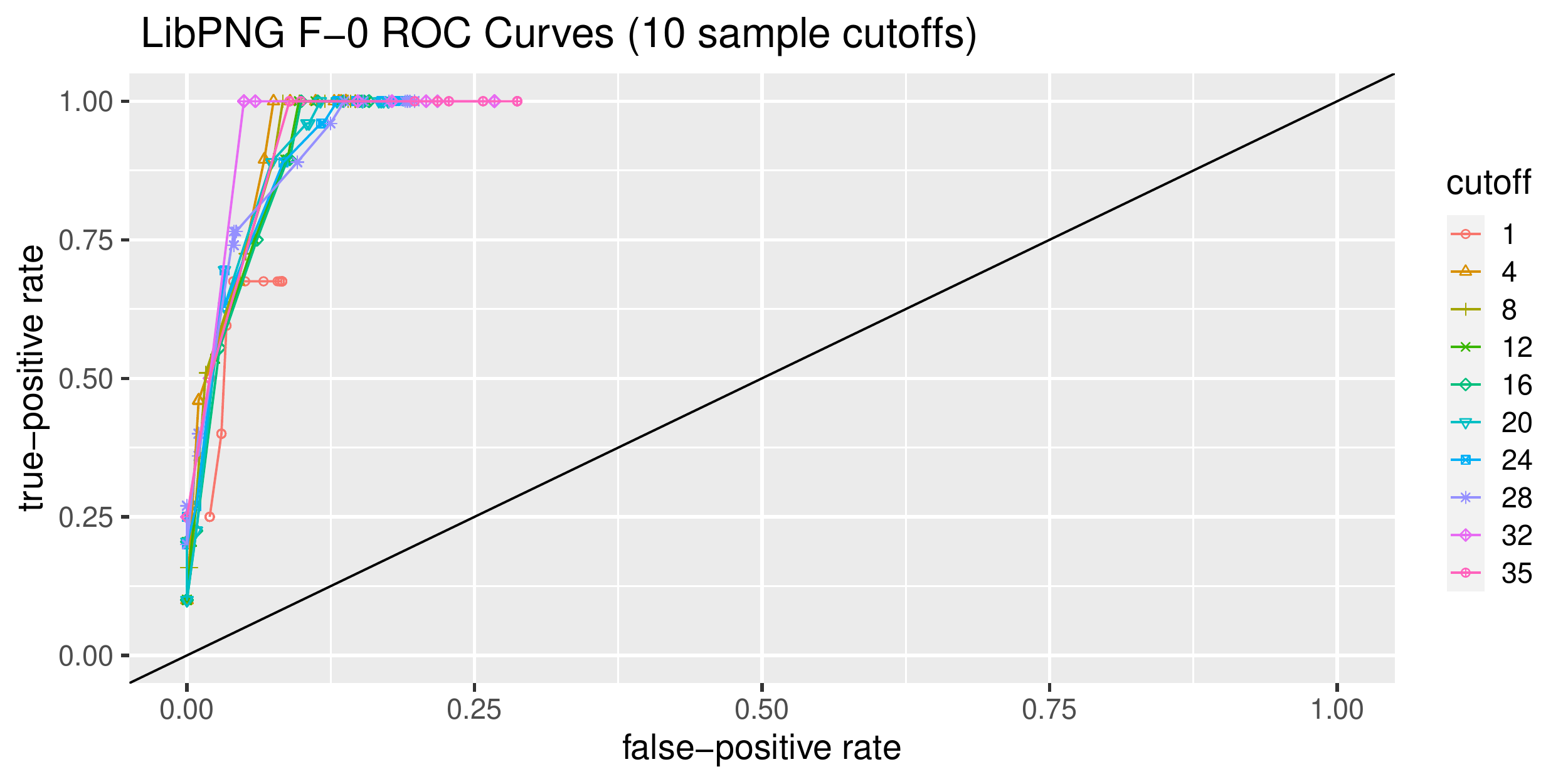} \\
  \caption{Impact of minimum score}
  \label{fig:min-compare}
\end{figure*}

\subsection{RQ3}

\noindent
Our third research question explores replacing \amodel's use of \LAVDNN with
\acount's frequency-based approach.
If the performance of \acount is similar, it suggests that \LAVDNN also
captures notions related to word frequencies.
Interestingly, if the alternative shows better performance, then it suggests
the use of dangerous words to augment the training of next generation of
predictors.
While the range of alternatives is virtually limitless, because we are
interested in the contribution of the terms, we consider a family of
algorithms that use \emph{term frequencies} to determine potentially dangerous words.  
For example, a straightforward algorithm would assert that all terms found in
vulnerable function names in the training data are dangerous.  
More sophisticated approaches would consider as negative evidence terms (frequently)
occurring in benign function names.

The family we consider increases a term's dangerousness when the term appears
in function names from the vulnerable training data and decreases it when the
term appears in names from the benign training data.
Thus, the algorithm assigns higher scores to terms that have high frequency
in the vulnerable training data and low frequency in the benign training data.
We refer to this as a \emph{family} because we consider %
\acount with a range of different \emph{weights} (a pair of the amount added
to, and the amount subtracted from, a term's score).

Algorithm~\ref{alg:favd}'s performance is impacted by the \cutoff and
\threshold values returned by \sfs{FindBest}.
This section first explores the \cutoff/\threshold search space.
It then considers a range of fixed weights, and finally considers an
algorithm that determines the best weight based on the training data.

\subsubsection{Exploration of \K and \threshold} 
\begin{figure}[b]
  \centering
  \includegraphics[width=0.99\columnwidth]{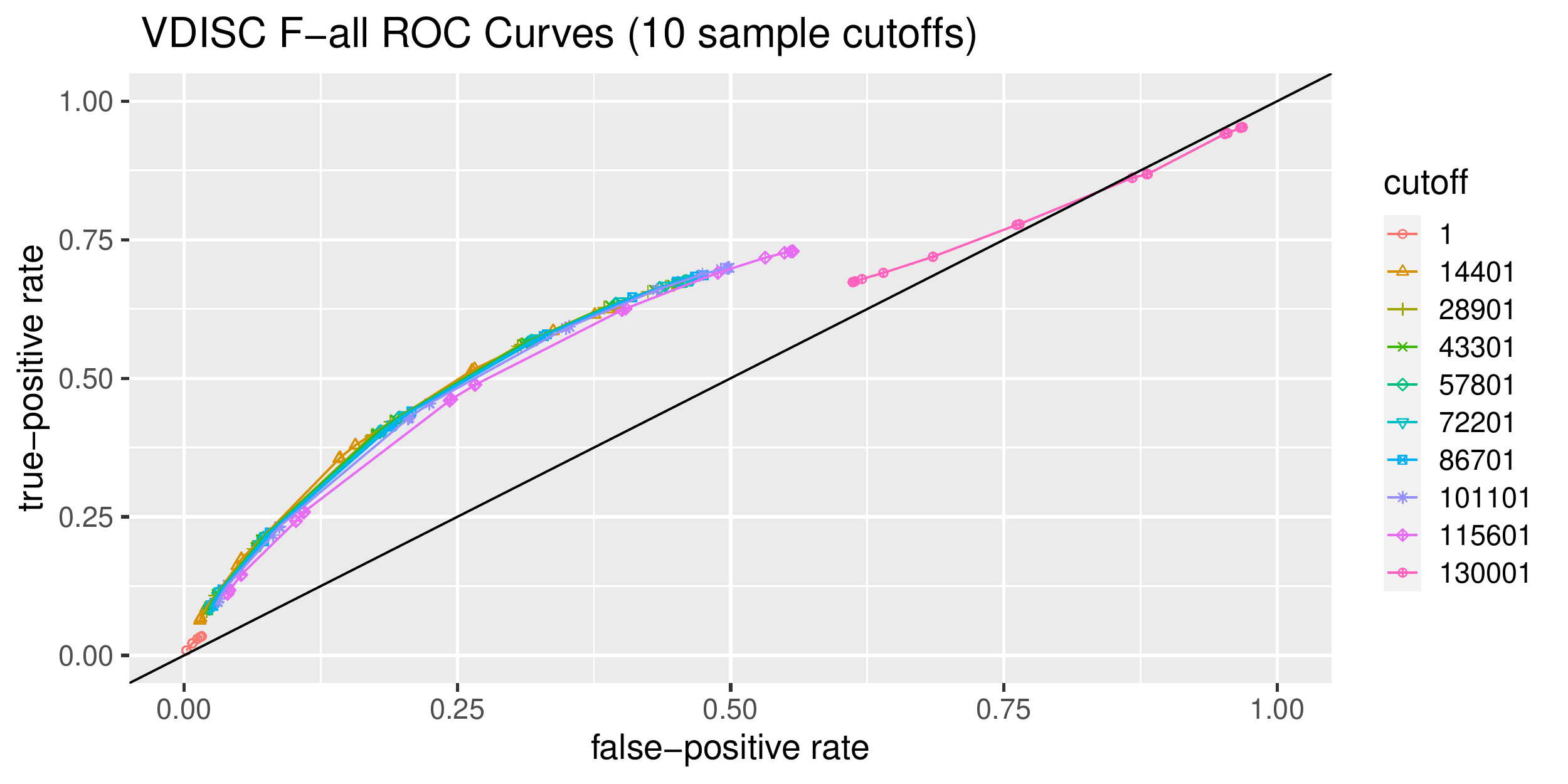} 
  \caption{ROC curves for \CWE}
  \label{fig:ROC-vdisc}
\end{figure}

The exploration starts by considering the
collection of ROC graphs
shown in Figures~\ref{fig:ROC-FFmpeg}--\ref{fig:ROC-vdisc}.
These graphs are for the optimal weights, which provide greater
discrimination and thus make visual patterns easier to observe.
A ROC graph plots the true-positive rate against the false-positive rate 
at various settings of a parameter (in our case \threshold). %
They are useful for comparing classifiers with each other and with a ``no
skill'' classifier.
An ideal model produces a ROC curve that goes straight up and then straight over to
the right, while the no-skill classifier produces a 45-degree line from the
origin to the upper right. 

When producing these graphs, we increment \cutoff in steps of 100 to reduce
visual clutter.
The \cutoff search landscape is reasonably smooth because, for example, in the
step from 4100 to 4200, the first 4100 words are the same.
Increments of 100 help speed up analysis and limit the size of the charts,
while having no meaningful impact on their interpretation.

The graphs help us understand the interplay between \K and \threshold.
When \threshold is 0.00, all function names are predicted to be vulnerable,
and thus, performance degenerates to that of the \strawman classifier.
At this point, both the true-positive rate and the false-negative rate are 1.00
and the ROC curve ends at the upper right of the chart.
{With the exception of Figure~\ref{fig:ROC-FFmpeg} we suppress this \threshold 
because it causes considerable visual clutter (explaining the absence of lines
to (1,1) in the other ROC charts).
}

To begin with, we consider the ROC graph for \sfs{FFmpeg} shown in 
\figref{fig:ROC-FFmpeg} where each ROC curve shows \threshold going from 1.00 to 0.00, 
while curve color shows \K going from small (red shift) to large (blue shift).
Increasing \K can be seen to have two effects:
first, it tends to flatten the curve, and second early on when \K goes from 1
to 101 the true-positive rate increases dramatically (from about 0.40 to about
0.65).
Here the true-positive rate increases much more than the false-positive rate.
However, by the time \K includes most of the words, the performance has degraded below
that of a no-skill classifier. 
Therefore, in practice, a user might choose to increase \cutoff until the
false-positive rate reaches some tolerance (i.e.,~patience for wrong answers).
Taken together, these two effects imply that the list of dangerous words
has its most dangerous words first.
Hence, we have our first interesting difference between \acount and \amodel,
because \amodel failed to effectively rank the dangerous words.

\figref{fig:min-compare} illustrates minimum score's impact.
These graphs show ROC curves for \sfs{LibPNG}, which is one of the datasets
where \acount excels.  
Hence, the ROC curves are all closer to the ideal ``up and over'' curve.
The ROC curves in these two charts show one of 10-sample \K's.
The upper chart in \figref{fig:min-compare} shows \call, which uses no minimum score, 
and thus includes all words as potentially dangerous words.
The lower chart shows \cz, where a minimum score of zero is used.
\cz's restriction to only high-scoring words clearly causes the ROC curves to
more closely resemble the ideal ``up and over'' curve than the flatter
(inferior) ROC curves of \call.
Finally, the comparison again indicates that the high-scoring words are more
likely to be associated with vulnerable functions.

\figref{fig:ROC-vdisc} shows the \CWE dataset (the \loo curves are similar).
The curves reinforce the pattern seen above where larger values of \K degrade
performance.
Because this dataset has the least useful vocabulary all \cutoff's except for
the smallest, 1, and the largest, 130\,001, show very similar overlapping
performance.
However, the patterns observed above are still evident.
For example, performance improves when using more of the vocabulary up until
the very end where, for the final \cutoff value, performance dips below that of
the no-skill classifier.

To summarize, there is a sweet-spot at
rather small values of \cutoff and \threshold.
Smaller \cutoff values include only the highest scoring words:
as seen in the ROC curves as \cutoff increases there is a ubiquitous flattening
of the curve.
Having a limited number of dangerous words works best when combined with a low
\threshold because this combination requires only a few of a function name's terms
to be on the dangerous words list.

\begin{figure}
  \centering
  \includegraphics[width=0.99\columnwidth]{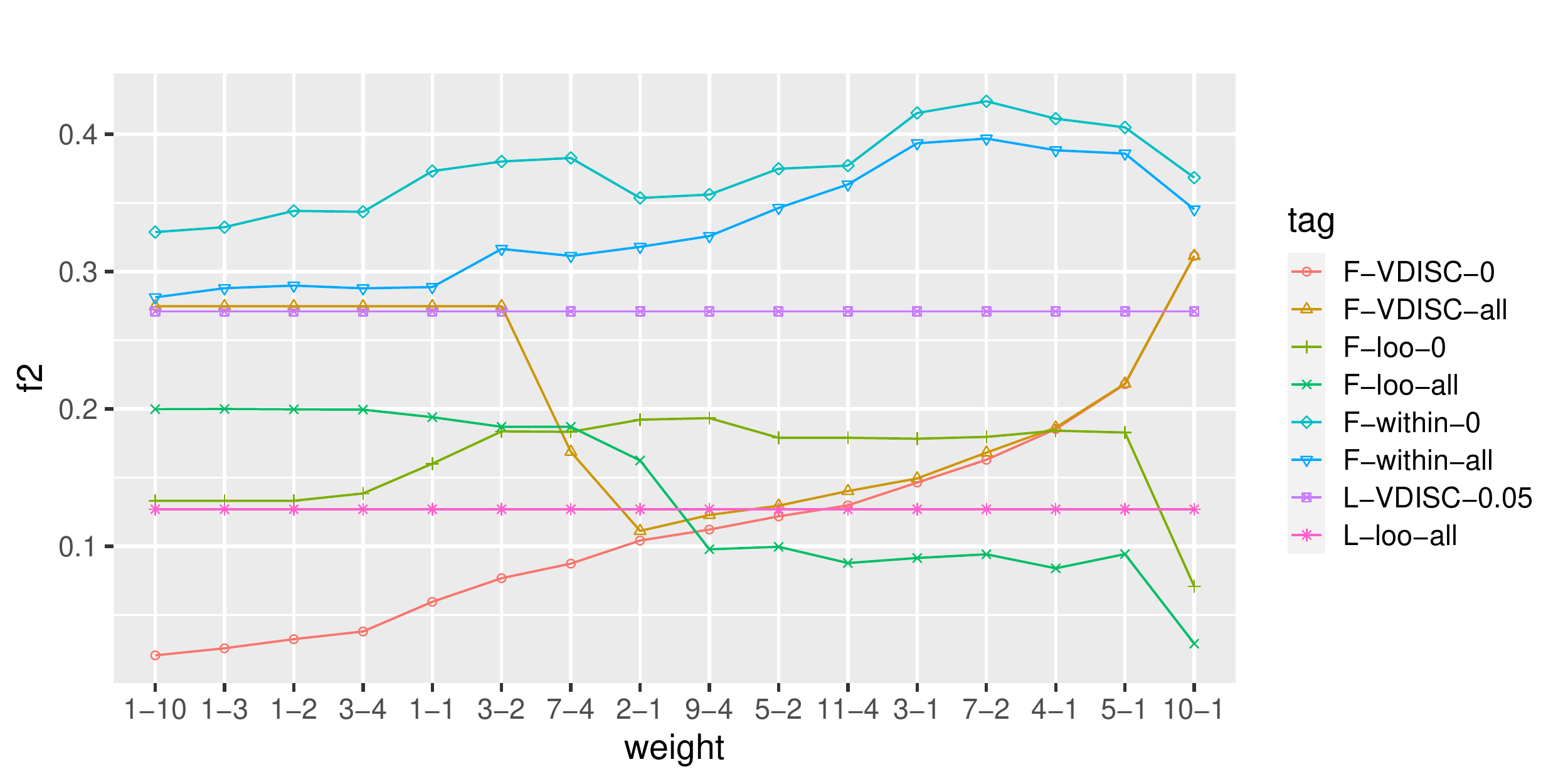}
  \caption{Comparison of \acount performance with fixed weights.}
  \label{fig:counting}
\end{figure}

\subsubsection{Exploring the impact of weight} 

\noindent
Initially, we explore the impact of a range of fixed weights, and later we
apply an algorithm that dynamically determines the best weight based on the
training data.
Figure~\ref{fig:counting} shows the performance of \acount for the three
datasets \win, \loo, and \CWE.
In this graph, the $x$-axis shows a range of \emph{weights}.
Each pair, \sfs{plus--minus}, shows the amount added to the score for each term
in a vulnerable function name followed by the amount subtracted for each
term in a benign function name.
For reference, the graph also includes the performance of \amodel,
which appears as a horizontal line because it is not affected by the weights.

Each line in the graph shows the performance of a specific ranking algorithm,
which we refer to using a \emph{tag} including three things: 
`\sfs{L}' for \amodel or `\sfs{F}' for \acount, the dataset involved: 
`\win`, `\loo', or `\vdisc', and the minimum score, where
`\sfs{all}' denotes that there is no minimum, and thus all words are included (in
which case the ranking affects only the order of the words).
For example, the tag \sfs{F-\win{}-0} is the top line, which applies \acount to
the \win dataset using a minimum score of 0.

Minimum score can help focus the analysis on high-scoring words.
This effect is evident in the two lines for \acount applied to the \win data
set, \sfs{F-\win{}-0} and \sfs{F-\win{}-all}.
It can also limit the available vocabulary, which can hurt performance.
This can be seen clearly in the case of \CWE where, on the left of the figure, 
\sfs{F-\vdisc-0} performs dramatically worse than \sfs{F-\vdisc-all}, then,
moving to the right as the weights favor inclusion, 
the performance gap disappears.
Thus, enforcing a minimum can help focus the algorithm on high-scoring words,
but it does so at the expense of limiting the available vocabulary.

Big picture, \acount performs best with the more focused vocabulary of the \win
dataset, which
shows the impact of having the right vocabulary.
In this case cross-validation \emph{within each project} clearly provides relevant vocabulary. 
The top two curves also illustrate the impact of applying a minimum score to the
list of dangerous words, which accounts for the gap between them.

For the \loo dataset, \acount struggles to outperform \amodel.  
When including all words, the performance clearly degrades moving from left to
right, which indicates that the list is loosing focus.
The implication here is that it is more important to reduce the importance of
terms from benign function names than to reinforce terms from vulnerable function names.
For example, with a weight of 1--10, a word has to be very rare in the benign data
to maintain a high score.
When using a minimum score of 0, this pattern is eventually seen at the far right.
However, at the far left the $\Fii$ score suffers because the minimum limits the
number of dangerous words to the detriment of the algorithm.
 This improves dramatically from \sfs{3-4} to \sfs{3-2} and levels off until it plummets at
the far right where unwanted words do not get enough negative weight.

Finally, for the \CWE dataset \acount is rarely able to outperform \amodel. 
It only truly does so at the far right where the weight finally concentrates
the truly dangerous vocabulary.  
Note that despite the positive slope on the right side of the graph, running
the weights out to \sfs{1000-1} shows no further improvement. 
The left of the graph parallels that of the \loo dataset with \sfs{F-\vdisc-0}
being starved for vocabulary.
One interesting feature of \sfs{F-\vdisc-all} is that to the left of \sfs{3-2},
the training data leads the algorithm to include all of the terms as potential
dangerous words.
Here the performance is similar to \sfs{L-\vdisc-0.05}, where the minimum
score of \sfs{0.05} includes all but the lowest-scoring words.

For the \sfs{F-\vdisc-all} data, it is interesting that the weights \sfs{1-10} and
\sfs{10-1} have similar performance (one might see \sfs{1-10} as all the
uninteresting words ``taking a step back'' while \sfs{10-1} as all the
interesting words ``taking a step forward'').
More formally, on the left for \sfs{F-\vdisc-0} only words that are absent from the benign list are
included because of the large \sfs{minus} value in the weights.
Using $V$ for vulnerable and $B$ for benign this is the set $V-B$.
At the far right the influence of $B$ is negligible and the resulting set is
effectively $V$.
Digging deeper and comparing the values of \cutoff used, 
for \sfs{1-10} each fold uses all of the 130 thousand unique words available.
In contrast, for \sfs{10-1}, only 14\% are used.
Thus, the important vocabulary is effectively concentrated by the weight
\sfs{10-1} better than the weight \sfs{1-10}.
That this concentration does not give notably better performance is
an indication of the lack of useful vocabulary in the \CWE dataset.
Therefore, we can percolate truly dangerous words to the beginning of the list, 
but lack a sufficient number of them to improve the $\Fii$ score.

\begin{table}\tablefont
\caption{Tukey's HSD for performance.}
\label{tab:group-performance}
\centering
\begin{tabular}{l@{}l l@{}r}
\toprule
\multicolumn{1}{c}{model-dataset-filter}
  & \multicolumn{1}{c}{$\Fii$}
  & \multicolumn{1}{c}{HSD} 
  & \multicolumn{1}{c}{DWC*} \\
  &&\multicolumn{1}{c}{group} & \multicolumn{1}{r}{average} \\
\midrule
\sfs{F-\win{}-0}    &   0.372   &   ~~~a      &            49     \\
\sfs{F-\win{}-all}  &   0.333   &   ~~~a      &      1$\,$613     \\
\sfs{L-\win{}-0.99} &   0.179   &   ~~~~~b    &           234     \\
\sfs{L-\win{}-0.90} &   0.172   &   ~~~~~b    &           462     \\
\sfs{L-\win{}-0.05} &   0.157   &   ~~~~~b    &      1$\,$081     \\
                 &           & {\footnotesize $(p < 0.0001)$}  &  \\

\midrule
\sfs{L-\loo{}-0.90} &   0.173   &   ~~~a      &      1$\,$898     \\
\sfs{F-\loo{}-0}    &   0.163   &   ~~~a      &           202     \\
\sfs{L-\loo{}-0.05} &   0.159   &   ~~~a      &      4$\,$374     \\
\sfs{L-\loo{}-0.99} &   0.154   &   ~~~a      &           976     \\
\sfs{F-\loo{}-all}  &   0.138   &   ~~~a      &      6$\,$618     \\
                 &           & {\footnotesize $(p = 0.9446)$}  &   \\

\midrule
\sfs{L-\vdisc-0.05} &   0.271   &   ~~~a     &     85$\,$383    \\
\sfs{L-\vdisc-0.90} &   0.246   &   ~~~a     &     38$\,$199    \\
\sfs{L-\vdisc-0.99} &   0.223   &   ~~~a     &     21$\,$905    \\
\sfs{F-\vdisc-all}  &   0.218   &   ~~~a     &    129$\,$966    \\
\sfs{F-\vdisc-0}    &   0.115   &   ~~~~~b   &     10$\,$782    \\
                 &           & {\footnotesize $(p < 0.0001)$}  &  \\
\bottomrule
\multicolumn{4}{l}{\footnotesize * DWC = Dangerous Word Count}
\end{tabular}
\end{table}

To objectively consider the vocabulary patterns, we statistically compare the
models used to construct Figure~\ref{fig:counting}.
  We also consider the impact of vocabulary size.
Table~\ref{tab:group-performance} summarizes the results of Tukey's Honestly
Significant Difference (HSD) applied to each dataset.
This test identifies specific treatment means that differ from each other
(those that do not share a letter).
For the \win dataset, \acount is clearly the more successful.
Of particular interest is that for \sfs{F-\win{}-0}, an average of only 49 words
are deemed dangerous, which is a mere 3\% of the 1613 words used by \sfs{F-\win{}-all}.
This is the frequency algorithm at its best.
The same pattern is seen with the \loo dataset where \sfs{F-\loo{}-0} selects
only 3\% of the dangerous words used by \sfs{F-\loo{}-all} (202 versus 6618).
However, none of the numeric differences for the \loo dataset are
statistically significant.
Still, it is interesting that \sfs{F-\loo{}-0} has better numeric performance while
using only 3\% of the dangerous words.
Finally, for \CWE, the opposite is true.
For example, comparing \sfs{F-\vdisc-all} and \sfs{F-\vdisc-0}, the use of a larger
vocabulary is accompanied by better performance.
The data in Table~\ref{tab:group-performance} reinforces the general pattern
where \acount shines when given a more focused vocabulary.

Finally, in the table, higher standard deviation is the reason that no statistically
significant differences are seen with the \loo dataset despite some of the
numeric differences in the $\Fii$ scores being on par with those attained using the
\CWE dataset.
This is not unexpected as the \CWE dataset is a large uniform dataset where,
as shown in Table~\ref{tab:data-sets}, the projects of \loo have very different
characteristics.
Summarizing the data shown in Table~\ref{tab:group-performance}, 
with the right well-focused vocabulary, \acount performs quite well indicating
that words valuable to the prediction exist.
The challenge can be finding this vocabulary.

\subsubsection{Determining the best weights}
\begin{figure}
  \centering
  \includegraphics[width=0.99\columnwidth]{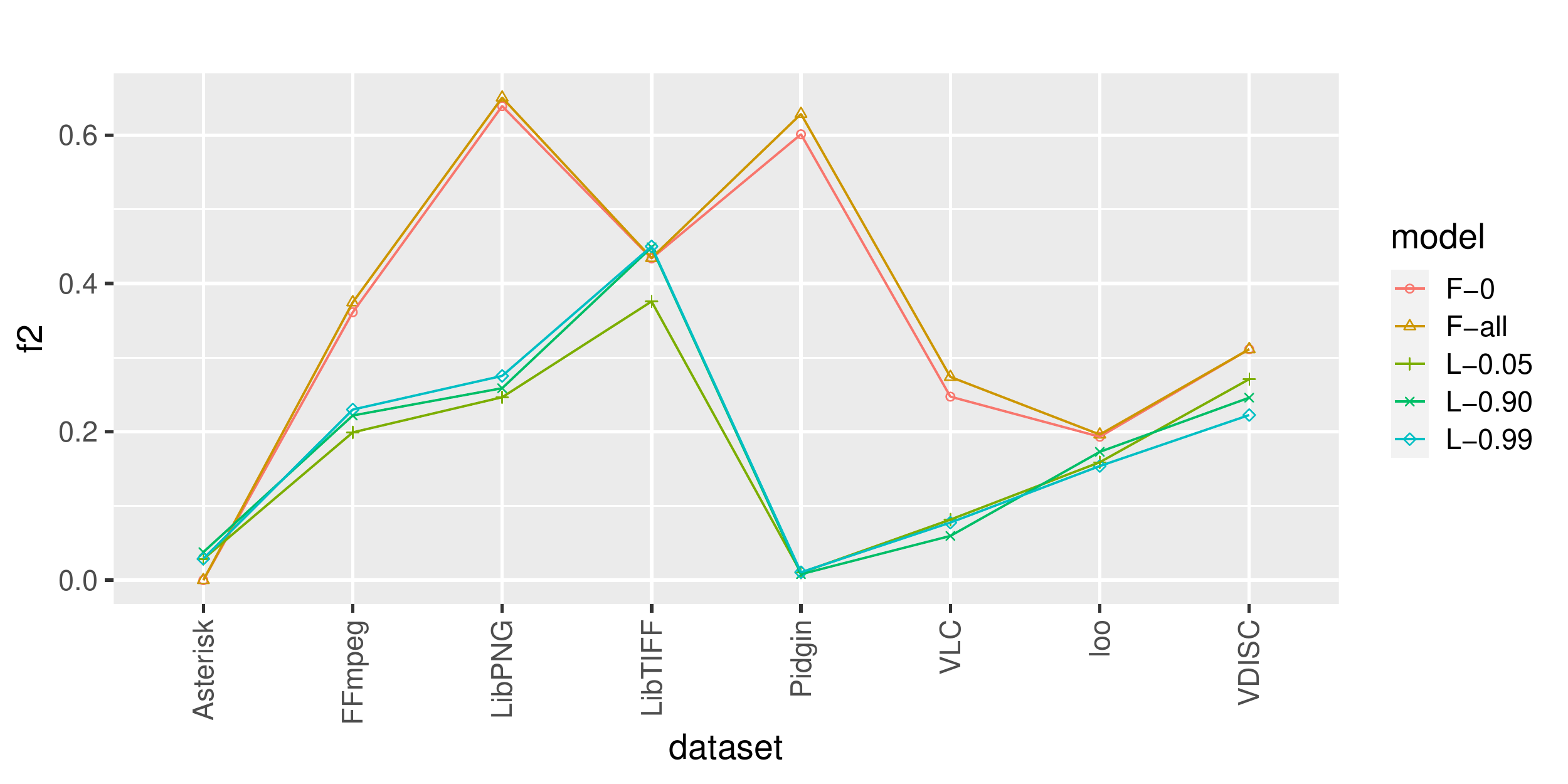}
  \caption{Dataset comparison (formally a bar graph is appropriate, but lines make the values visually easier to compare).}
  \label{fig:rq3-ds-cmp}
\end{figure}

In production, \acount determines the best weight based on the training data.
Our current algorithm uses a simple brute-force search through the list of
weights used to create in Figure~\ref{fig:counting}, and selects the weight
providing the best performance. 
The result is shown in Figure~\ref{fig:rq3-ds-cmp}, which compares \acount using all words and a minimum score of zero with the best performing
minimum scores for \amodel for the \win, \loo, and \CWE datasets (0.99, 0.90,
and 0.05, respectively).
The $x$-axis shows the dataset while the $y$-axis shows the $\Fii$ score.
Note that in the following, we look at the individual datasets of \win separately.

From an interpretability viewpoint, there is clear evidence in
Figure~\ref{fig:rq3-ds-cmp} that \LAVDNN is not exploiting terms to the extent
possible.
Specifically, \sfs{Pidgin}, and to a lesser extent \sfs{LibPNG}, showcase
\acount's advantage over \amodel at exploiting a focused vocabulary.
While less pronounced, the same is true of \sfs{FFmpeg} and \sfs{VLC}.
Of the remaining two projects from the \win dataset \sfs{Asterisk} proves
universally hard to predict while \sfs{LibTIFF} is comparatively easy to
predicted for both \amodel and \acount.

Table~\ref{tab:ds-comparison} shows the results of ANOVAs separately
comparing the two \acount models (\call and \cz) with the three top-performing
\amodel models.
\acount's performance shows that terms have unexploited value in six of the
nine datasets where \acount performs better than \amodel.
Furthermore, its performance is inferior on only one (where the $p$-value of
0.0264 is not a strong endorsement).
These results reinforce the general pattern seen in the previous analysis, 
where the frequency models excel when using smaller and less diverse vocabularies.
The only \win dataset that \acount truly struggles with is \sfs{Asterisk},
which, as seen in Table~\ref{tab:data-sets}, has the largest vocabulary and
lowest percentage of vulnerable functions.

\begin{table}[b]\tablefont
\caption{Comparing \acount with the maximum \amodel performance (\textbf{bold} shows statistically significant improvement).}
\label{tab:ds-comparison}
\centering
\raisebox{-33mm}{\rotatebox{90}{\textsl{minimum score 0 (\cz)}\hspace*{7mm}\textsl{all words (\call)}}}
\setlength{\tabcolsep}{2pt}
\begin{tabular}{l@{~~}c c r c}
\toprule
\multicolumn{1}{c}{dataset}
  & \multicolumn{1}{c}{\acount $\Fii$}
  & \multicolumn{1}{c}{max(\amodel $\Fii$)} 
  & \multicolumn{1}{c}{p-value} 
  & \multicolumn{1}{c}{limit(\acount $\Fii$)} \\
\midrule
\sfs{overall}        &  \textbf{0.355} & \textrm{0.182} &  $< 0.0001$   &  \textrm{0.442}  \\
\sfs{Asterisk}       &  \textrm{0.000} & \textbf{0.038} &  $  0.0264$   &  \textrm{0.060}  \\
\sfs{FFmpeg}         &  \textbf{0.375} & \textrm{0.230} &  $< 0.0001$   &  \textrm{0.386}  \\
\sfs{LibPNG}         &  \textbf{0.651} & \textrm{0.275} &  $  0.0003$   &  \textrm{0.850}  \\
\sfs{LibTIFF}        &  \textrm{0.450} & \textrm{0.450} &  $  0.7046$   &  \textrm{0.542}  \\
\sfs{Pidgin}         &  \textbf{0.629} & \textrm{0.011} &  $< 0.0001$   &  \textrm{0.667}  \\
\sfs{VLC}            &  \textbf{0.247} & \textrm{0.082} &  $  0.0111$   &  \textrm{0.410}  \\
\loo                 &  \textrm{0.196} & \textrm{0.173} &  $  0.9767$   &  \textrm{0.314}  \\
\CWE                 &  \textbf{0.311} & \textrm{0.271} &  $< 0.0001$   &  \textrm{0.330}  \\
\midrule
\sfs{overall}        &  \textbf{0.345} & \textrm{0.182} &  $< 0.0001$   &  \textrm{0.416}  \\
\sfs{Asterisk}       &  \textrm{0.000} & \textbf{0.038} &  $  0.0264$   &  \textrm{0.000}  \\
\sfs{FFmpeg}         &  \textbf{0.361} & \textrm{0.230} &  $< 0.0001$   &  \textrm{0.381}  \\
\sfs{LibPNG}         &  \textbf{0.639} & \textrm{0.275} &  $  0.0004$   &  \textrm{0.868}  \\
\sfs{LibTIFF}        &  \textrm{0.434} & \textrm{0.450} &  $  0.7051$   &  \textrm{0.503}  \\
\sfs{Pidgin}         &  \textbf{0.601} & \textrm{0.011} &  $< 0.0001$   &  \textrm{0.678}  \\
\sfs{VLC}            &  \textbf{0.247} & \textrm{0.082} &  $  0.0419$   &  \textrm{0.388}  \\
\loo                 &  \textrm{0.193} & \textrm{0.173} &  $  0.9814$   &  \textrm{0.221}  \\
\CWE                 &  \textbf{0.311} & \textrm{0.271} &  $< 0.0001$   &  \textrm{0.330}  \\
\bottomrule
\end{tabular}
\end{table}

While it is easy to get drawn into the relative comparison of the $\Fii$
scores, we are also interested in the relative contribution that terms make
to the discrimination of vulnerable functions.
{
To provide an indication of how much room there is for improvement, the last
column of Table~\ref{tab:ds-comparison} shows the best $\Fii$ score that \acount 
attains on the training data.
While not a hard limit on its performance with the test data, typically, for
the given list of dangerous words, the $\Fii$ score on the training data
provides an upper bound for a value seen using the test data.  
Thus, this column provides an indication of the best that one might
expect to attain using only the terms found in the function names. 
}
While some projects such as \sfs{LibPNG} and \sfs{LibTIFF} show room for
improvement, projects such as \sfs{Pidgin} and \sfs{FFmpeg} are within 0.038
and 0.011 of their maximum $\Fii$ score. 

Finally, Table~\ref{tab:dwcc} shows the number of dangerous words used by \acount for
each dataset found on the $x$-axis of Figure~\ref{fig:rq3-ds-cmp}.
What is quite striking in the table is how well the small vocabularies
perform when using a minimum score of zero (e.g.,~with \sfs{LibPNG} and \sfs{LibTIFF}).
While the smaller vocabularies never produce a numerically higher $\Fii$ score,
\emph{none} of the differences is statistical significant (the smallest
$p$-value is 0.729).

\begin{table}\tablefont
\centering
\caption{Comparison of dangerous words counts. 
}
\label{tab:dwcc}
\setlength{\tabcolsep}{6pt}
\begin{tabular}{lrrrrr}
\toprule
&\multicolumn{2}{c}{$\Fii$ score}  
  &\multicolumn{3}{c}{dangerous words count} \\
\multicolumn{1}{c}{dataset}
  &  \multicolumn{1}{c}{min 0}
  & \multicolumn{1}{c}{all}
  & \multicolumn{1}{r}{min 0}
  & \multicolumn{1}{r}{all} 
  & \multicolumn{1}{c}{percent} \\ 
\midrule
\sfs{Asterisk} &   0.000   &   0.000   &     22  &   3184  &  0.70 \%  \\ 
\sfs{FFmpeg}   &   0.361   &   0.375   &    126  &   2111  &  5.96 \%  \\ 
\sfs{LibPNG}   &   0.639   &   0.651   &     31  &    327  &  9.42 \%  \\ 
\sfs{LibTIFF}  &   0.434   &   0.434   &     61  &    434  & 14.11 \%  \\ 
\sfs{Pidgin}   &   0.601   &   0.629   &     28  &   1886  &  1.47 \%  \\ 
\sfs{VLC}      &   0.247   &   0.274   &     38  &   1738  &  2.16 \%  \\ 
\loo           &   0.193   &   0.196   &    213  &   6618  &  3.22 \%  \\ 
\CWE           &   0.311   &   0.311   &  18478  & 129966  & 14.22 \%  \\ 

\bottomrule
\end{tabular}
\end{table}

\subsubsection{Summary} 
Returning to RQ3's ``Does direct construction of the dangerous words list provide insight into \LAVDNN?''
The answer is a resounding ``yes.'' As seen in Table~\ref{tab:ds-comparison},
\acount outperforms \amodel for most datasets.
While on an absolute scale the $\Fii$ values are not high, the important take
home message here is that there is utility in the terms that \LAVDNN is failing to
exploit.
Hence, the training of future Deep Neural Networks aimed at vulnerability
prediction should include features based on the terms in the hope of better
exploiting their potential.
Furthermore, the relative performance on \win where the vocabulary is more
focused, suggests two additional things.
First, considered in the context of an evolving system where past data from the
same system can be used in the prediction, \acount alone provides a
lightweight pre-filter to activities such as manual code review.
The second interesting implication comes from \acount attaining respectable
performance on some of the datasets (i.e.,~\sfs{Pidgin}'s $\Fii$ score of 0.651
and \sfs{LibPNG} of 0.629). 
Specifically, these results are achieved using vanishingly little training data
from the neural net training perspective (e.g.,~only 26 and 31 vulnerable
function names exist in the data for \sfs{Pidgin} and \sfs{LibPNG},
respectively).

\subsection{Threats To Validity}

\noindent
We identify potential threats to the validity of our experimental design and evaluation.
One such threat is that the source of our \CWE dataset is a collection of functions extracted from open-source projects~\cite{russell2018:automated} 
that are unknown to us, both in scope, demographics, and domain.
This may have resulted in an unknown bias effect regarding our results for \CWE. 

Furthermore, we evaluate on only six project-based datasets.
It would mitigate this threat to external validity if the number of such datasets was increased.  
We continue to look for additional datasets to use in our experiments.
The external validity of our diversity observations could also be improved by
access to more data with a known heterogeneity.

Moreover, the greedy search used by \sfs{FindBest} trades precision (finding
a global maximum) for speed. %
Thus, we prioritize performance, but internal validity could improve using
techniques with a lower chance of getting stuck in local maxima.
While this search is an approximation because local maxima exist, exhaustive
searches using the smaller datasets found the error was only a few percent.      

A further threat arises from the tools used in our study. 
There may be defects in the implementation
that escaped our testing, thereby affecting our results.
The same is true of \LAVDNN, which was provided by its authors.

Finally, the statistical tests used are all well established and their
implementations publicly available in \sfs{R}, and thus well vetted.
However, it is possible that more appropriate tests unknown to us might provide
more appropriate evidence.
We also endeavored to follow the most up-to-date information from the
statistics community when interpreting the models~\cite{wasserstein2019:moving}.

\section{Related Work}
\label{sec:relwork}

\noindent
A source code vulnerability is a weakness in the source code that can be exploited into a security issue. 
Publicly known vulnerabilities are organized by common identifiers in the
Common Vulnerabilities and Exposures (CVE) database~\cite{mitre:cve}, 
where they are classified using the Common Weakness Enumeration (CWE)~\cite{mitre:cwe}, 
and ranked using the Common Vulnerability Scoring System (CVSS)~\cite{first:common}.

Over the last two decades, various methods have been presented to identify potential security 
vulnerabilities in code based on static program analysis~\cite{pistoia2007:survey,kulenovic2014:survey}.
The recent advances and successes in machine learning (ML) have resulted 
in an increased interest in adapting these techniques to the vulnerability prediction problem~\cite{pang2017:predicting,
ghaffarian2017:software,handa2019:machine,li2019:comparative,jiang2019:survey}.
However, the choice of feature types, classifiers, and data balancing techniques 
has a large impact on %
the prediction's performance~\cite{kaya2019:impact}

The \emph{naturalness hypothesis}~\cite{allamanis2018:survey} states that source code exhibits 
similar statistical properties as other forms of human communication. 
This means that corpus-based statistical learning can capture the local regularity in source code,
i.e.,~such models can predict with high accuracy what code to expect in a given context, 
or what properties such code should have. 
One application of this idea is the use of deep feature representation learning on lexed C and C++ source code 
for automatic function-level vulnerability detection~\cite{russell2018:automated}. 

Recurrent Neural Networks (RNN) and Long Short Term Memory networks (LSTM) 
have been successfully applied for code reviews and vulnerability detection~\cite{gupta2018:intelligent,fan2019:software,xu2019:vulnerability},
for example by training a Bidirectional LSTM on so-called code gadgets, 
which are collections of semantically related lines~\cite{li2018:vuldeepecker}.
Semantic properties of code can be predicted using code2vec~\cite{alon2019:code2vec},
which represents code snippets as a fixed-length code vector, 
very similar to how word2vec~\cite{mikolov2013:distributed} represents sentences.
Harer et al.~find that ML-based vulnerability prediction trained directly on C and C++ source code performs better 
than alternative approaches that were trained using semantic (build-time) information for the same source code 
(such as control flow information and def-use relations)~\cite{harer2018:automated}. 

One of the main challenges to all of these approaches is that they are
computationally expensive to develop, as well as to keep 
up to date with newly discovered vulnerability patterns.
The \LAVDNN research~\cite{li2019:lightweight} that inspired this paper is 
an example of a more lightweight approach with more modest goals.
By using function names as semantic cues for training a Deep Neural Network (DNN), 
the model aims to predict potential vulnerability of a function based on its name, 
with the goal of helping a developer (or analysis technique) focus on 
those functions that should be scrutinized more carefully.

\section{Concluding Remarks}
\label{sec:conc}

\noindent
\LAVDNN~\cite{li2019:lightweight} attempts lightweight
function vulnerability prediction based solely on the function's name. 
This paper takes that idea a step further and explores feather weight
prediction based solely on the terms that make up function names.
In doing so, this paper aims to provide an interpretability viewpoint
for better understanding of an otherwise ``black box'' DNN. %
{We find that \LAVDNN has limited ability to identify dangerous words,
and generally can not provide an effective ranking of those words.}

Comparing the relative performance of \FAVDL and \FAVDF allows us to probe whether
\LAVDNN learns ``dangerous words'' or attempts classification in other ways.
Generally, \FAVDL provides much
more consistent (though not very good) performance. 
When faced with a diverse dataset such as \CWE, the performance is generally
better than \FAVDF.
However, with the more focused vocabulary of a single system, \FAVDF
dramatically outperforms \FAVDL. From this, we can conclude that \LAVDNN
has some ability to identify dangerous words, but does \emph{not} generally
provide an effective ranking of those words.

The performance difference suggests that there is room to augment DNN's with
networks that directly learn dangerous words, particularly in more mature
projects with more stable vocabularies.
It is also worth noting that there are many cases in which \FAVDF outperforms
the publicly available implementation of \LAVDNN. 
This improvement suggests two things.
First, augmentation with a dangerous words predictor can
improve predictions if appropriate context is detected.
The second implication is that for a mature project, \FAVDF might replace \LAVDNN as an
even lighter weight predictor.
In this case, \FAVDF has the advantage that it needs orders of magnitude
less training data.

Future work includes qualitative analysis that uses the ground truth to assess
to what extent the vulnerability predictions by the two approaches overlap, are
in conflict, or are complementary. 
When \LAVDNN is to be augmented by other word finding algorithms, 
it is critical to understand what words \LAVDNN already finds.
{Another possibility is to try to determine if the DNN picks up on
patterns involving various n-grams (e.g.,~3-grams composed of three consecutive
letters from an identifier) and also on non-adjacent letter patterns in the
identifiers.}
With respect to the technique itself, one area for future work is to consider 
alternative sources of dangerous words.
These need not all be code-based.
For example, issues noted during requirements solicitation might provide a
source of additional dangerous words. 

As a possible \LAVDNN augmentation, we plan to investigate heavier-weight vocabulary possibilities.
For example, we plan to train a similar model to \LAVDNN on our own data, 
and investigate the value of splitting identifiers during the DNN model training phase. 
We also plan to consider sources of vocabulary beyond function names to improve the prediction, 
such as, formal parameters, called functions, and alike.

\section*{Acknowledgements}

\noindent
Dr.\ Moonen's work is supported by the Research Council of Norway through the secureIT project (RCN contract \#288787).

\balance

\printbibliography

\end{document}